\documentstyle[12pt,epsf]{article}
\newcommand{\postscript}[2] 
{\setlength{\epsfxsize}{#2\hsize}  
\centerline{\epsfbox{#1}}}
\newcommand{\nc}{\newcommand}
\nc{\fdiag}{0}
\nc{\bg}{B. Grzadkowski}
\nc{\BG}{Bohdan Grzadkowski}
\nc{\lsp}{\;\;\;\;\;\;\;\;}
\nc{\beq}{\begin{equation}}   \nc{\eeq}{\end{equation}}
\nc{\bea}{\begin{eqnarray}}   \nc{\eea}{\end{eqnarray}}
\nc{\baa}{\begin{array}}      \nc{\eaa}{\end{array}}
\nc{\bit}{\begin{itemize}}    \nc{\eit}{\end{itemize}}
\nc{\ben}{\begin{enumerate}}  \nc{\een}{\end{enumerate}}
\nc{\bce}{\begin{center}}     \nc{\ece}{\end{center}}
\nc{\non}{\nonumber}
\nc{\lumun}{\;{\hbox {fb}^{-1}}{\hbox {yr}^{-1}}}
\nc{\hc}{\hbox {h.c.}}
\nc{\re}{\hbox {Re}}
\nc{\im}{\hbox {Im}}
\nc{\etal}{\hbox{et al.}}
\nc{\pbarn}{\;\hbox {pb}}
\nc{\prd}[3]{{\it Phys.\ Rev.}\ {{\bf D{#1}} (#2), #3}}
\nc{\prl}[3]{{\it Phys.\ Rev.\ Lett.}\ {{\bf {#1}} (#2), #3}}
\nc{\plb}[3]{{\it Phys.\ Lett.}\ {{\bf B{#1}} (#2), #3}}
\nc{\npb}[3]{{\it Nucl.\ Phys.}\ {{\bf B{#1}} (#2), #3}}
\nc{\ptp}[3]{{\it Prog.\ Theor.\ Phys.}\ {{\bf {#1}} (#2), #3}}
\nc{\zfp}[3]{{\it Z.\ Phys.}\ {{\bf C{#1}} (#2), #3}}
\nc{\mpla}[3]{{\it Mod.\ Phys.\ Lett.}\ {{\bf A{#1}} (#2), #3}}
\nc{\rmp}[3]{{\it Rev.\ Mod.\ Phys.}\ {{\bf {#1}} (#2), #3}}
\nc{\ijmpa}[3]{{\it Int.\ J.\ of\ Mod.\ Phys.}\
               {{\bf A{#1}} (#2), #3}}
\nc{\app}[3]{{\it Acta\ Phys.\ Polon}\ {{\bf B{#1}} (#2), #3}}
\nc{\ra} {\rightarrow}
\nc{\ctw}{\cos\theta_W} 
\nc{\stw}{\sin\theta_W}
\nc{\ctwsq}{\cos^2\theta_W}        
\nc{\stwsq}{\sin^2\theta_W}
\nc{\ttbar}{t\bar{t}}
\nc{\bbbar}{b\bar{b}}
\nc{\tanb} {\tan \beta}
\nc{\twbdec} {t\rightarrow W^+ b}
\nc{\tbwbdec} {\bar{t} \rightarrow W^- \bar{b}}
\nc{\hprod} {e^+e^- \ra Z^\ast \ra H Z}
\nc{\epem} {e^+e^-}
\nc{\wpwm} {W^+W^-}
\nc{\tbar} {\bar{t}}
\nc{\bbar} {\bar{b}}
\nc{\wpp} {W^+}
\nc{\mt}{m_t}
\nc{\mts}{m_t^2}
\nc{\mw} {m_W}
\nc{\mws} {m_W^2}
\nc{\mz} {m_Z}
\nc{\mzs} {m_Z^2}
\nc{\mh} {m_H}
\nc{\mhs} {m_H^2}
\nc{\ma} {m_A}
\nc{\mas} {m_A^2}
\nc{\hdec}{H \ra t\bar{t}}
\nc{\ttbardec}{\ttbar \ra W^+W^-\bbbar}
\nc{\po}{\Phi_1}
\nc{\pod}{\Phi_1^\dagger}
\nc{\pht}{\Phi_2}
\nc{\phtd}{\Phi_2^\dagger}
\nc{\phtt}{{\tilde{\Phi}}_2}
\nc{\popo}{\po^\dagger\po}
\nc{\phtpt}{\pht^\dagger\pht}
\nc{\popt}{\po^\dagger\pht}
\nc{\phtpo}{\pht^\dagger\po}
\nc{\sq}{\sqrt{2}}
\nc{\nsd} {N_{SD}}
\nc{\ntt} {N_{tt}}
\nc{\vs}{\vspace{2mm}}
\nc{\sty}{\hat{S}^t_1} \nc{\pty}{\hat{P}^t_1}
\nc{\sts}{(\sty)^2}      \nc{\pts}{(\pty)^2}
\nc{\yts}{\sts+\pts}
\nc{\sby}{\hat{S}^b_1} \nc{\pby}{\hat{P}^b_1}
\nc{\sbs}{(\sby)^2}      \nc{\pbs}{(\pby)^2}
\nc{\ybs}{\sbs+\pbs}
\nc{\eettz}{\epem \rta \ttbar Z}

\newcommand{\zgvepjzgae}{(g^v_{Vee}+j g^a_{Vee})}
\newcommand{\zvepjzae}{(g^v_{Zee}+j g^a_{Zee})}
\newcommand{\zgat}{g^a_{Vtt}}
\newcommand{\zgvt}{g^v_{Vtt}}
\newcommand{\zgaf}{g^a_{Vff}}
\newcommand{\zgvf}{g^v_{Vff}}
\newcommand{\zat}{ g^a_{Ztt}}
\newcommand{\zvt}{ g^v_{Ztt}}
\newcommand{\ImPih}{\im{\Pi_h(p_h^2)}}
\newcommand{\RePih}{\re{\Pi_h(p_h^2)}}
\newcommand{\Pipropkmz}{\Pi_Z(s)}
\newcommand{\Pipropkmzg}{\Pi_V(s)}
\newcommand{\Piqpqb}{\Pi_{t+\bar{t}}}
\newcommand{\Piqmqb}{\Pi_{t-\bar{t}}}
\newcommand{\mEpsilon}{\epsilon(p_-,p_+,p_t,p_{\bar{t}})}
\newcommand{\mzg}{m_V}
\nc{\et}{\tilde{e}}
\nc{\ft}{\tilde{f}}
\nc{\gt}{\tilde{g}}
\nc{\hti}{\tilde{h}}

\def\lsim{\mathrel{\raise.3ex\hbox{$<$\kern-.75em\lower1ex\hbox{$\sim$}}}}
\def\gsim{\mathrel{\raise.3ex\hbox{$>$\kern-.75em\lower1ex\hbox{$\sim$}}}}

\def\fbi{~{\rm fb}^{-1}y^{-1}}
\def\fbinver{~{\rm fb}^{-1}}

\def\gev{\,{\rm GeV}}
\def\tev{\,{\rm TeV}}

\def\rta{\rightarrow}

\def\mib#1{\mbox{\boldmath $#1$}}

\begin{document}
%
\font\fortssbx=cmssbx10 scaled \magstep2
\medskip
\begin{flushright}
$\vcenter{
\hbox{\bf IFT-7/99}
\hbox{\bf UCD-99-9} 
\hbox{June, 1999}
}$
\end{flushright}
\vspace*{1.5cm}
\begin{center}
{\large{\bf Testing Top-Quark Yukawa Interactions in $\!\!\mib{e}^+\mib{e}^-
\!\to\mib{t}\bar{\mib{t}}\mib{Z}$}}\\ 
\rm
\vspace*{1cm}

{\bf \BG}
\footnote{E-mail:{\tt bohdang@fuw.edu.pl}} {\bf and}
{\bf Jacek Pliszka}
\footnote{E-mail:{\tt pliszka@fuw.edu.pl}}\\ 

\vspace*{1cm}
{\em Institute of Theoretical Physics, Warsaw University, 
Warsaw, Poland\\
and\\
Davis Institute for High Energy Physics, 
UC Davis, CA, USA }\\

\vspace*{2cm}

{\bf Abstract}
\end{center}
\vspace{5mm} 

Determination of the top-quark Yukawa couplings in the process 
$e^+e^- \rightarrow t \bar{t} Z$ has been studied for the high luminosity
option ($\int L=500\fbinver$) of a linear $e^+e^-$ collider. Polarization of the
electron beam has been considered. The method of optimal 
observables has been adopted to determine the couplings and to disentangle
different models. It has been found that it could be possible to discriminate
between the Standard Model scalar Higgs boson, pure pseudoscalar and the
Higgs boson of mixed CP at  
about $12 \sigma$ level for $\sqrt{s}=1.\tev$ for the Higgs boson
mass slightly above twice the top-quark mass both for the leptonic and hadronic
top-quark decay channels. However, $\sqrt{s}=.5\tev$ collider turned out to
have too 
small energy for tests of the top-quark Yukawa coupling because of its
limited production 
rate (at most $2.5 \sigma$ effect). Possible tests of CP violation in the
scalar sector have been considered.

\vfill
\setcounter{page}{0}
\thispagestyle{empty}
\newpage

\renewcommand{\thefootnote}{\sharp\arabic{footnote}}
\setcounter{footnote}{0}

\section{Introduction}
It will be very important to directly determine the Yukawa couplings (and
therefore CP nature) of any Higgs
boson that would be discovered. Although it is very plausible
that a substantial number of Higgs boson events will first be
available at the LHC, however future linear $e^+e^-$ colliders (NLC) 
could provide much more efficient laboratory to determine couplings of 
any observed Higgs.
The Standard Model (SM) predicts an existence of a scalar Higgs boson,
however a pseudoscalar admixture is absolutely conceivable and as leading to
renormalizable in\-ter\-actions, 
it may couple as strongly as the scalar Higgs
boson. Since Yukawa couplings are proportional to the appropriate fermion
masses, obviously the most promising coupling to study is the one of the
top quark.
Here, we are going to discuss a possibility to test the top-quark Yukawa 
coupling in the process $\eettz$, our goal will be to estimate what
would be the future precision for discrimination between CP-conserving
(scalar and pseudoscalar) and CP-violating Yukawa interactions.

For our analysis, the relevant part of the interaction Lagrangian 
takes the following form:

\beq
{\cal{L}}= -\frac{\mt}{v}h\tbar(a+i\gamma_5 b)t + 
c \frac{g \mz}{2\ctw} h Z_\mu Z^\mu,
\label{lag}
\eeq
where $g$ is the $SU(2)$ coupling constant, 
$v$ is the Higgs boson vacuum expectation value 
(with the normalization adopted here such that $v=2m_W/g=246\,\gev$) 
$a$, $b$ and $c$ are parameters which account for possible deviations from the
SM ($a=1$, $b=0$ and $c=1$ reproduce the SM Lagrangian). 
Since under CP $\tbar(a+i\gamma_5 b)t \stackrel{CP}{\ra} \tbar(a-i\gamma_5
b)t$ and $Z_\mu Z^\mu 
\stackrel{CP}{\ra} Z^\mu Z_\mu$ one can observe that terms in the cross
section proportional to 
$ab$ or $bc$ would indicate CP violation. Since for the process considered
here ($\eettz$) there are no $ab$~\footnote{Those terms could appear
for polarized $\ttbar$.} terms 
we will focus here on $bc$ terms. It should be
emphasized that this is not just $bc$ term but also $(bc)^2$ which are
signals of CP violation.

The minimal extension of the SM which provides non-zero $b$ and $a,c\neq 1$
is the two-Higgs-doublet model~\cite{weinberg} (2HDM). 
Since the model contains two SU(2) Higgs doublets it could be
shown~\cite{weinberg} that CP violation may arise in the 
Higgs sector. Such CP violation is explicit if there is no choice of phases
such that all the Higgs-potential parameters are real. However,
even if all potential parameters can be chosen to be real, spontaneous
CP violation (complex vacuum expectation values) is a possibility. 
Therefore, the 2HDM 
is a very attractive and simple model in which to explore
the implications of CP violation in the Higgs sector. 
After SU(2)$\times$U(1) gauge symmetry breaking, 
if either explicit or spontaneous CP violation is present,
the three neutral degrees of freedom (one corresponding to the Goldstone boson
decouples) mix in the mass matrix and as a result of the mixing between 
real and imaginary parts of neutral
Higgs fields, the Yukawa interactions of the mass-eigenstates are not
invariant under CP. They are given exactly by form of the Lagrangian
defined in Eq.(\ref{lag}). Because of the mixing, the form of the $ZZh$
coupling will also be modified by the factor $c$. Both $a$, $b$ and $c$ 
are calculable functions of the mixing angles. 

Recently, CP violation in a scalar sector has attracted much more attention as it
has been realized that it may be generated within 
the Minimal Supersymmetric Standard 
Model (MSSM),
where due to the CP-violating soft-supersymmetry-breaking Lagrangian
complex phases emerge in the effective potential generated
at the one-loop level of the perturbation expansion~\cite{PilWag}.
It turns out that the mixing between the pseudoscalar $A$ and the heavier
Higgs boson $H$  can be as large as
$25\%$ leading to substantial modification of $ZZh_1$ and $Zh_1h_2$, where
$h_{1,2}$ are the physical mass-eigenstates. Another consequence of the mixing
would be CP-violating Yukawa interaction of the form given in
Eq.(\ref{lag}). 

However, in this paper we are not restricting ourselves to any 
particular model, on the contrary, 
our analysis is supposed to be as model independent as possible
and we will treat the Lagrangian given by Eq.(\ref{lag}) as an effective interaction.

The process $\eettz$ has been already discussed in the literature by
Hagiwara~\cite{hagiwara} in the context of the SM. Since the total luminosity
for the $\epem$ linear collider considered there was a factor of $50$ lower 
than
the recently discussed in the context of the TESLA project~\cite{tesla} we
found it was worth to reconsider the process applying some more
elaborate statistical methods to test the Yukawa couplings in the
framework of the effective Lagrangian without referring to any given model. 
Bar-Shalom, Atwood and Soni first noticed in their paper~\cite{Atwood-Soni} that the
process $\eettz$ could be used to test CP violation in the Higgs sector,
the advantage of the process is that CP violation appears even at the
tree-level approximation of the perturbation expansion. This is why its
consequences could be relatively large. However Ref.~\cite{Atwood-Soni}
restricts its discussion to the specific case of the 2HDM 
what makes its applicability much more
limited. We shall compare our results both with Ref.~\cite{hagiwara} and
Ref.~\cite{Atwood-Soni}. 

The paper is organized as follows. In Section 2 we will briefly review the
method of the optimal observables. Section 3 will contain a description of
our strategy and results for the determination of the top-quark Yukawa
couplings. Section 4 will be devoted to the summary and conclusions concerning
tests of the top-quark Yukawa couplings at the NLC. In the Appendix, 
an analytic form for the contribution
to the matrix element squared linear in the CP-violating couplings will be
presented. 

\section{The Optimal Analysis Procedure}

We shall consider the process of $\ttbar Z$ production at linear $\epem$
colliders:
\beq
e^+(p_+)+e^-(p_-) \ra t(p_t) + \tbar(p_{\tbar}) + Z(k),
\eeq
where momenta of the particles involved have been indicated. Since the
electron beam could be relatively easily polarized, besides unpolarized
beam we will discuss electrons with positive and negative helicity.
The differential cross section for the process could be
written in the following way:
\beq
\Sigma(\phi)\equiv {d\sigma \over d\phi}
=\left[f_{VV}+ac f_{ac} +(ac)^2 f_{(ac)^2}+bc f_{bc}+(bc)^2 f_{(bc)^2}
\right],
\label{sigform}
\eeq
where $f_{VV}$ stands for non-Higgs contribution, whereas other $f_i$ are
defined through coefficients standing in front of them. $\phi$ denotes the final
state phase space configuration, 
$c$ is the factor which parameterize the $ZZh$ coupling,
$a$ and $b$ are the scalar- and pseudoscalar-type Yukawa
couplings defined in Eq.(\ref{lag}).

Here, we wish to discriminate between various models 
on the basis of the di\-ffer\-ence 
between the phase space distribution
dependence of $f_{bc}$, $f_{ac}$, $f_{(bc)^2}$ and $f_{(ac)^2}$.

In general, for $\Sigma(\phi)=\sum_i c_if_i(\phi)$
(where $f_i(\phi)$ are known functions, 
including normalization)
the determination of the unknown signal coefficients $c_i$ 
with the smallest possible statistical error is given~\cite{oo} by 
\beq
c_i=\sum_k M_{ik}^{-1}I_k\,,\quad{\rm where}~~
M_{ik}\equiv \int {f_i(\phi)f_k(\phi)\over \Sigma(\phi)} d\phi\,,~{\rm and}~
I_k\equiv \int f_k(\phi) d\phi\,.
\label{ciform} 
\eeq
The covariance matrix for the $c_i$ is
\begin{equation}
V_{ij}\equiv \langle \Delta c_i\Delta c_j\rangle= 
{ M_{ij}^{-1} \sigma_T\over N}=M_{ij}^{-1}/L_{\rm eff}\,,
\label{cerror}
\end{equation}
where $\sigma_T=\int {d\sigma\over d\phi} d\phi$ is the integrated
cross section and $N=L_{\rm eff}\sigma_T$ is the total number of events,
with $L_{\rm eff}$ being the luminosity times the efficiency.
This result is the optimal one regardless of the
relative magnitudes of the different contributions
to $\Sigma(\phi)$. 
It is equivalent to determine $c_i$ by maximizing the
likelihood (in the Gaussian statistics limit) of the fit to the full $\phi$
distribution of all the events. 
The increase in errors due to statistical fluctuations in
the presence of background is implicit in the possible
background contribution to $\Sigma(\phi)$ appearing in Eq.~(\ref{ciform}),
which implies larger 
$M_{ij}^{-1}$ entries in Eq.~(\ref{cerror}).  From the result 
of Eq.~(\ref{cerror}), 
the $\chi^2$ in the $c_i$ parameter space is then computed as 
\beq
\chi_X^2(Y)=\sum_{i,j}(c_i^\prime-c_i)V_{ij}^{-1}(c^\prime_j-c_j)=
\sum_{i,j}(c^\prime_i-c_i)L_{\rm eff}M_{ij}(c^\prime_j-c_j)\,,
\label{chisqform}
\eeq
where, for our theoretical
analyses, the $c_i$ and $V_{ij}$ are 
computed within the model $X$, while $c_i^\prime$ should be calculated for
the model $Y$  tested against the model $X$. The value of $\chi_X^2(Y)$
indicates how well could one distinguish $X$ and $Y$.
{\bf The power of the optimal analysis technique is to take
full advantage of all the information
available in the cross section as a function of the kinematical variables.}

The above procedure is not altered if cuts are imposed
on the kinematical phase space over which one integrates; one simply restricts
all $\phi$ integrals to the accepted region.
If a subset, $\bar\phi$, of the kinematical
variables $\phi$ cannot be determined, then the optimal technique 
can be applied using the variables, $\hat\phi$, that {\it can} be observed
and the functions $\bar f_i(\hat\phi)\equiv \int f_i(\phi) d\bar\phi$.
Since cuts restrict an information available in a given process, one should
expect that the sensitivity of optimal observables would be reduced after
imposing cuts.

\section{Strategy and Results}

\begin{figure}[h]
\postscript{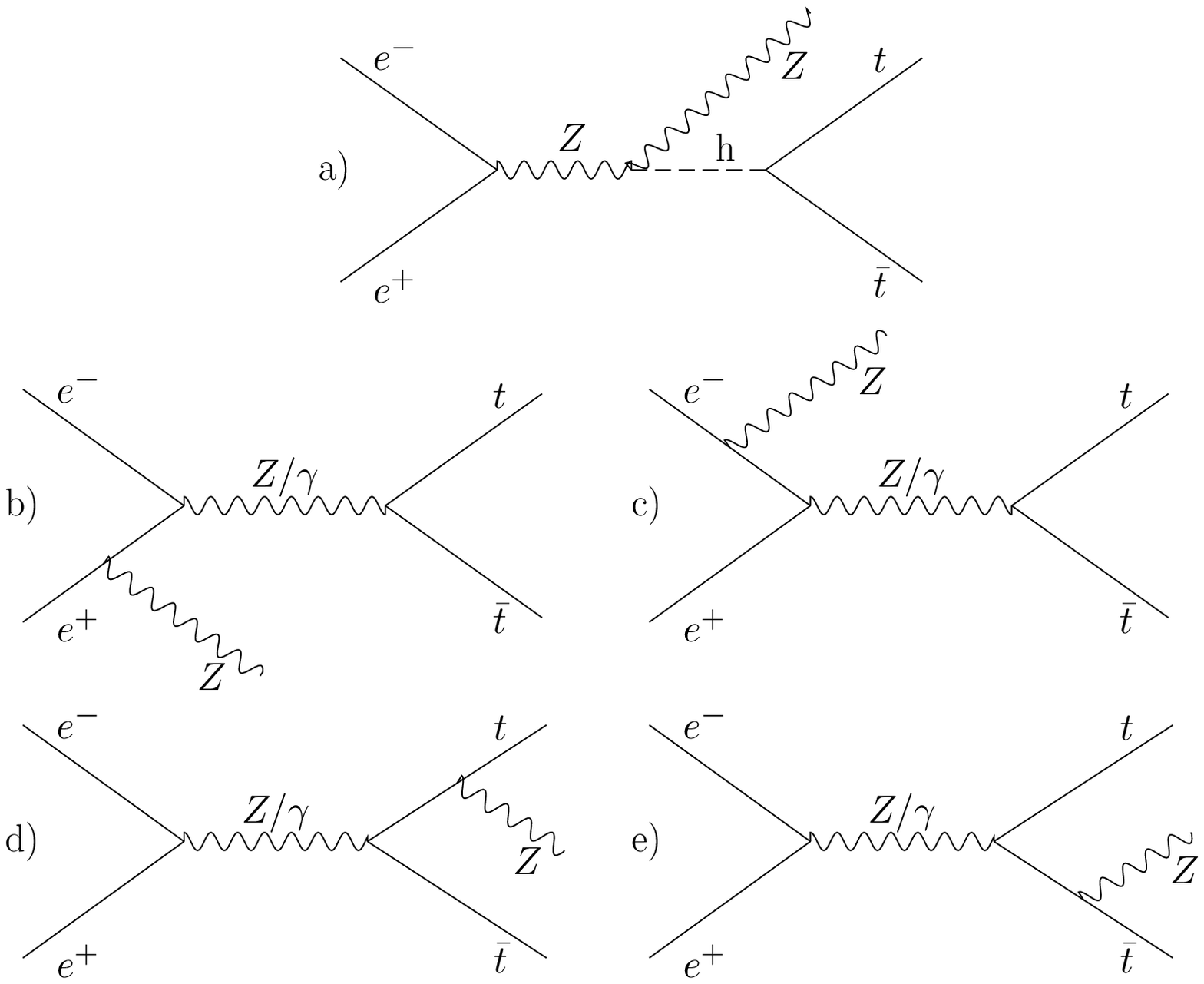}{.8}
\vspace*{-1.2cm}
\caption{The Feynman diagrams for the process $\eettz$.}
\label{diagrams}
\end{figure}

Feynman diagrams describing the process $\eettz$ are shown in
Fig.~\ref{diagrams}. 
One can think about the process considered here as a way to measure
the top-quark Yukawa interaction of a Higgs boson produced in a strahlung
process off the $Z$ line $\epem \ra Zh$ followed by the decay $h \ra \ttbar$.
However, here we allow also for Higgs boson being off-shell, see the
diagram (a) in Fig.~\ref{diagrams}. Within this interpretation the remaining
(purely SM-type) diagrams (b), (c), (d) and (e) describe a continuum
background for the Higgs boson decaying to $\ttbar$ pair. 
Since we are trying to determine the top-quark Yukawa coupling
the contribution to the matrix element squared relevant for us would be the
interference of the Higgs-boson-exchange diagram (a) with 
$Z$-boson radiation off electron line (b), (c) and top-quark line (d), (e)
and the square of the Higgs 
diagram (a). In some sense the process 
$\eettz$ considered here is complementary to $\epem \ra \ttbar h$ discussed
recently in Ref.\cite{GGK}, as both processes originate from the
Higgs-boson strahlung off $Z$ line followed either by $h \ra \ttbar$ or
$Z \ra \ttbar$, respectively.

\begin{figure}[h]
\vspace*{-1.5cm}
\postscript{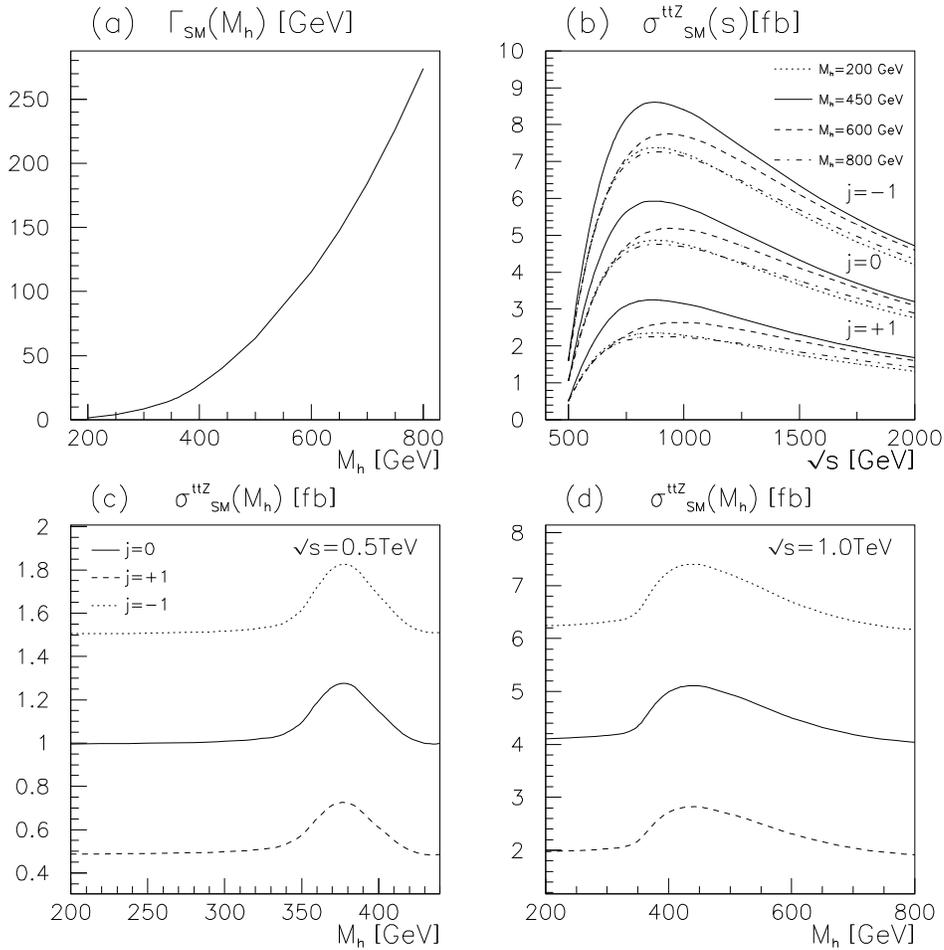}{.8}
\vspace*{-2.8cm}
\caption{The SM predictions for the total Higgs-boson width (a), the total cross section
for $\eettz$ for indicated polarization of the initial electron beam
($j=-1,0,+1$) as a function of $\protect\sqrt{s}$ for fixed
Higgs-boson mass $M_h$ (b), and the total cross section as a function of $M_h$ for 
$\protect\sqrt{s}=.5\tev$ (c) and for $\protect\sqrt{s}=1.\tev$ (d).}
\label{cross}
\end{figure}

\begin{figure}[h]
\vspace*{-2.7cm}
\postscript{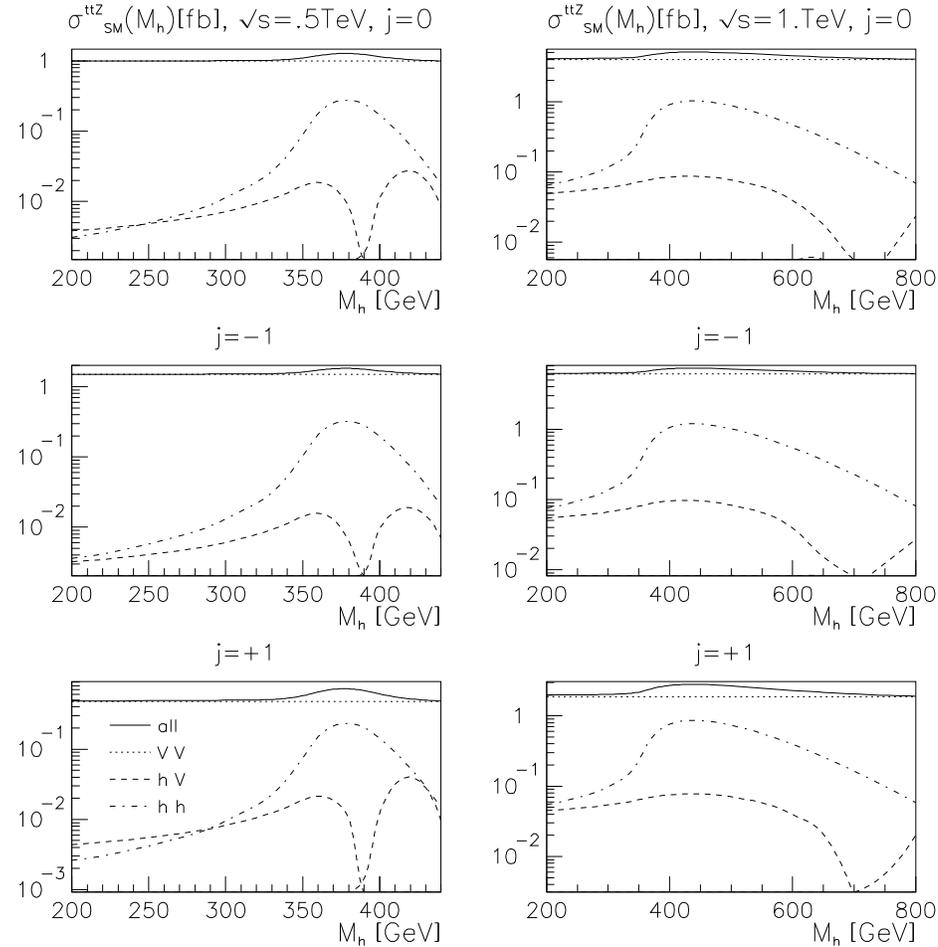}{.8}
\vspace*{-2.6cm}
\caption{The SM predictions for the total cross section
for $\eettz$ for indicated polarization of the initial electron beam
($j=-1,0,+1$) as a function of Higgs-boson mass $M_h$ for
$\protect\sqrt{s}=.5\tev$ and for $\protect\sqrt{s}=1.\tev$.
Contributions from the Higgs-boson diagram squared (hh), all non-Higgs
diagrams (VV), and from the interference
between Higgs and non-Higgs diagrams (hV) have been plotted separately.
The interference becomes negative for $M_h>390\gev$ and $700\gev$ at
$\protect\sqrt{s}=.5\tev$ and $\protect\sqrt{s}=1.\tev$, respectively.
Therefore we have plotted the absolute value of the interference term.   
The solid line stands for the sum of all the contributions.}
\label{interfer}
\end{figure}

Hereafter we will be using the top-quark mass $\mt=175\gev$, $Z$-boson mass
$\mz=91.17\gev$ and the Weinberg angle $\sin^2 \theta=0.23$. Electrons will
be considered as massless.
Before we start the optimal analysis it is useful to recollect the SM predictions
for the total Higgs-boson width $\Gamma_h^{SM}$ and also the total cross
section for the process of interest ($\eettz$) for various $\sqrt{s}$ and
Higgs-boson masses $M_h$, for polarized $j=\pm 1$ and unpolarized $j=0$
electron beams. Results are presented in Fig.~\ref{cross}~\footnote{Using
our code we have obtained exactly the results presented for the SM by Hagiwara
$\etal$ in Ref.~\cite{hagiwara}. Other test we have performed was to
reproduce the narrow-width-approximation for the on-shell Higgs boson. 
We have also checked our code against the analytical form shown in the Appendix for the
contribution linear in $bc$.}. Because the Higgs 
boson could be on its mass shell, the
Breit-Wigner form of the propagator must be adopted, therefore it is
relevant to remind ourself the
strong increase of $\Gamma_h$ as a function of $M_h$, see plot (a) in Fig.~\ref{cross}.
Concerning the total cross section as a function of $\sqrt{s}$ we observe a
threshold increase at $\sqrt{s}=\mz+2\mt$ followed by a typical s-channel
suppression for higher energies. For the polarization dependence, the
relative suppression of the polarization $j=1$ could
have been anticipated, as it is caused by a constructive and destructive
interference between $\gamma$ and $Z$ exchange diagrams for $j=-1$ and
$j=+1$, respectively.
As it is seen from plots (c) and (d) in Fig.~\ref{cross} the relative
contribution from the Higgs boson to the total 
cross section could reach up to $30\%$ in the region where $2\mt < M_h <
\sqrt{s} - \mz$.
It could be read from Fig.~\ref{cross} that for a given $\sqrt{s}$ the
total cross section 
for low Higgs-boson mass, $M_h=200$, is greater than the one for high mass.
That means that the Higgs-boson-exchange diagram (a) adds constructively
with the other diagrams for low masses and interferes destructively in the
region of high masses. In fact, it is seen from Fig.~\ref{interfer} where
separate contributions to $\sigma_{tot}(\eettz)$ have been plotted.
This is related to the correct high-energy
behavior of the cross section in the large $M_h$ limit. 
Since the virtual
$Z$ boson couples to massless electrons, effectively it is mostly
the transverse component which propagates and therefore the Higgs-boson
exchange is not necessary to provide ``good'' high-energy behavior
in the infinite Higgs-boson mass limit.~\footnote{For a detailed discussion
of unitarity constraints see Ref.~\cite{hagiwara}.}.
Quantitatively however, the effect is totally negligible for the total
cross section. 

Following detector design of the Japanese Linear Collider~\cite{JLC} we
impose here the polar angle cut $|\cos\theta| < 0.98$. We will consider
separately two cases. For the first one, which we call the leptonic case,
one top quark decays hadronically and the other leptonically
and the second one, which we call the hadronic case in which 
both $t$ and $\tbar$ decay hadronically. We choose the following $\ttbar$ tagging
efficiencies:
$.30$ for leptonic and $.44$ for hadronic modes. For hadronic decays it is
very difficult to disentangle experimentally $t$ and $\tbar$, we would be
only able to tell from the invariant mass reconstruction that a given jet
corresponds to $t$ or $\tbar$. Therefore, adopting the optimal observables
for hadronic $\ttbar$ decays, we will always be averaging over $t$ and $\tbar$
that would correspond to $t$ and $\tbar$ tagging just by the invariant mass
measurement without determining the charge of decaying top quarks. Obviously,
for leptonic decays it is not necessary since the charge of
hadronically decaying top quark would be identified by the lepton charge
from the semileptonic decay of the other top quark. For 
simplicity we will assume $100\%$ efficiency for $Z$ tagging. 
For definiteness, in what follows, we will consider one year of operation 
for high luminosity option of the NLC that has been examined in the context
of the TESLA collider design, for which one expects $L=500\fbi$.~\cite{tesla}.

\newpage

We first focus on the detection of CP violation in the process $\eettz$. As
has already been mentioned CP, violation would result in the appearance of
non-zero $bc$ and $(bc)^2$ terms in the differential distribution,
Eq.(\ref{sigform}). 
An interference of the Higgs-exchange Feynman diagram with the $Z$ radiation off
the final top-quark lines
generates terms $bc$, whereas
$(bc)^2$ are emerging from the square of the Higgs-exchange diagram.  
Let us assume that the non-standard CP-odd admixture to 
the Yukawa coupling
is small, so that the terms proportional to $(bc)^2$ could be neglected and
the leading contribution is the one linear in $bc$. In the Appendix we present the
analytical formula for the contributions to the matrix element squared
that are proportional to $bc$. It turns out that in the small $bc$ limit the differential
cross section could be written in the form $\Sigma(\phi)=\Sigma_i c_i
f_i(\phi)$ and the optimal analysis outlined in the previous section could
be directly applicable~\footnote{Notice that the Higgs-boson width
$\Gamma_h$ enters the matrix element squared through the denominator of the
Higgs boson 
propagator. Since $\Gamma_h$ depends on $a^2$, $b^2$ and $c^2$ the differential
cross section is not a polynomial in couplings to be determined, therefore
in general 
Eqs.(\ref{ciform},\ref{cerror}) are not applicable. However, for small $b$
one can expand the propagator around its SM form. Since we are
going to keep only linear terms in $b$, effectively we are allowed to use
the SM form of the propagator while calculating the functions $f_i$.
Therefore, in the small $b$ limit, the 
differential cross section is 
of the desired form indeed.}. 

\begin{figure}[h]
\vspace*{-1.5cm}
\postscript{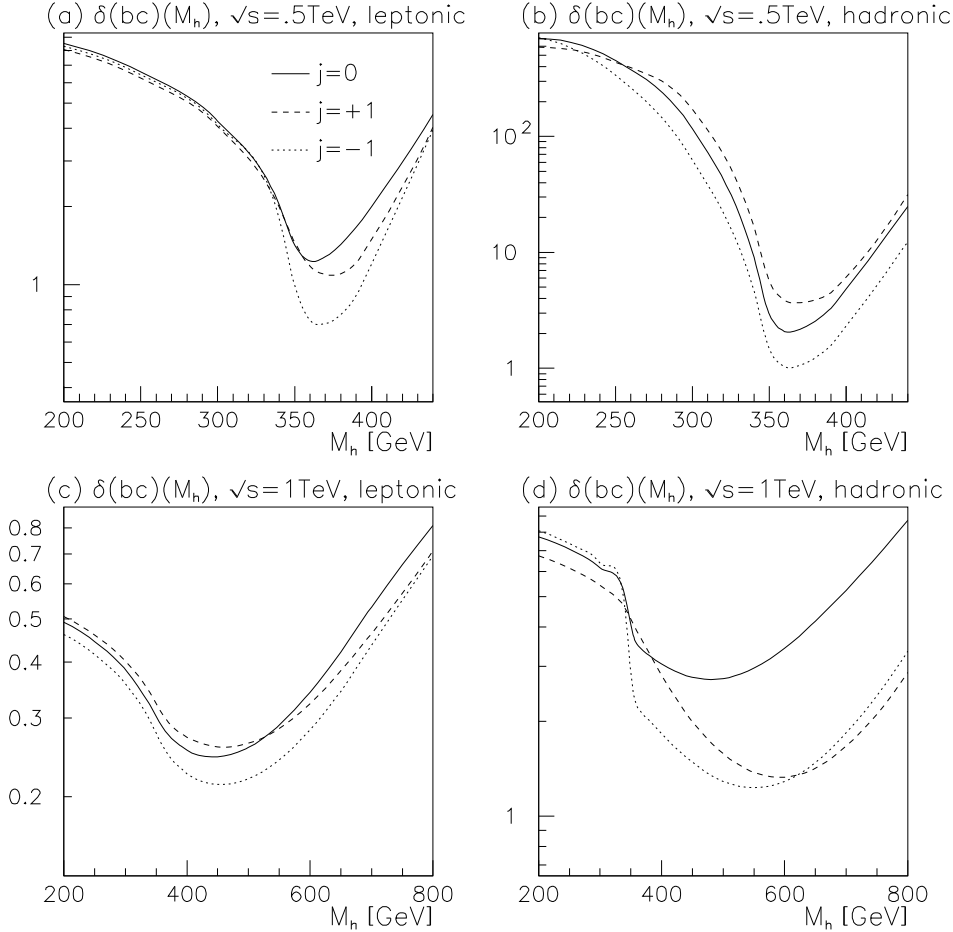}{.8}
\vspace*{-2.8cm}
\caption{The statistical error $\delta(bc)$ for the determination of $bc$ for the case when
$(bc)^2$ could be neglected, calculated for the indicated polarizations of the
initial electron beam and $\protect\sqrt{s}=.5,\;1.\tev$.}
\label{smallbc}
\end{figure}

As it has been shown in Ref.~\cite{oo} one can
construct an appropriate weighting function $\omega_i(\phi)$ such that
$\int \omega_i(\phi) \Sigma(\phi)=c_i$. The covariance matrix for such
observables is given by the Eq.(\ref{cerror}). In our simple case when only
linear terms in the unknown coupling $bc$ are kept, the weighting function for
determination of $bc$ and its statistical error become:
\bea
\omega_{bc}&=&\left[\int\frac{f_{bc}^2}{\Sigma_{SM}}\right]^{-1}
\frac{f_{bc}}{\Sigma_{SM}} \\
\label{omega}
\delta(bc)&=&\left[\frac{1}{N}\frac{\int \Sigma_{SM}}{\int
\frac{f_{bc}^2}{\Sigma_{SM}}}\right]^{1/2},
\label{bcerror}
\eea
where $\Sigma_{SM}$ denotes SM contribution to the differential cross
section. Numerical results for $\delta(bc)$ are shown in Fig.~\ref{smallbc}.
As seen from the figure, the smallest errors appear for $M_h$ such
that the Higgs boson could be on its mass shell and decay to $\ttbar$, this
is the region corresponding to the maximal cross section, see
Fig.~\ref{cross}. In general, hadronic top-quark decay modes provide much
larger errors, however in the vicinity of 
$M_h=370\gev$ both hadronic and leptonic final states lead to errors 
of the order of $1.$ for $\sqrt{s}=.5\tev$. For 
$\sqrt{s}=1.\tev$ minimal errors are obtained for $M_h=450\gev$ for
leptonic decays and one can
see that leptonic modes offer $\delta(bc)\simeq .25$ whereas for the
hadronic ones we get $\delta(bc) > 1.$. The enhancement of the error for the
hadronic modes illustrates how important the $t-\tbar$ identification is.
As expected, loosing information on the top-quark charge leads to
larger errors.

Let us now consider the general case with non-negligible terms ${\cal
O}(b^2)$. Then the 
dependence of the differential distribution Eq.({\ref{sigform}) on the
unknown couplings $a$, $b$ and $c$ is much more involved since the
Higgs-boson width, which enters the denominator of the Higgs-boson
propagator also depends 
on those couplings. In other words, functions $f_{i}$
appearing in the distribution shown in Eq.(\ref{sigform}) are unknown as
being $a$, $b$ and $c$ dependent.
However, we shall assume here that at the time when the NLC will operate the total
Higgs-boson width $\Gamma_h$ will be known from measurements at the LHC.
Therefore we are going to 
consider only those extensions to the SM that predict the same $\Gamma_h$,
otherwise the model would be experimentally excluded at the very beginning. 
Moreover, the measurement of the total cross section for the process $\eettz$
would also 
eliminate those models for which the cross section would differ from the
measured one. Therefore, in addition, we will restrict ourself to those models which
provide the total cross section identical with the SM one~\footnote{The
choice of the SM values for the width and the cross section is, of course,
arbitrary. It serves just a possible example of the future experimental
data.}. The models we are discussing are specified by the couplings $a$, $b$ and
$c$ appearing in the Lagrangian given by Eq.(\ref{lag}), the following
versions will be considered:
\bit
\item SM: $a=1$, $b=0$, $c=1$
\item CP-mix.1: $a=b>0$
\item CP-mix.2: $a=b<0$
\item CP-odd: $a=0$
\eit

\begin{figure}[t]
\vspace*{-1.5cm}
\postscript{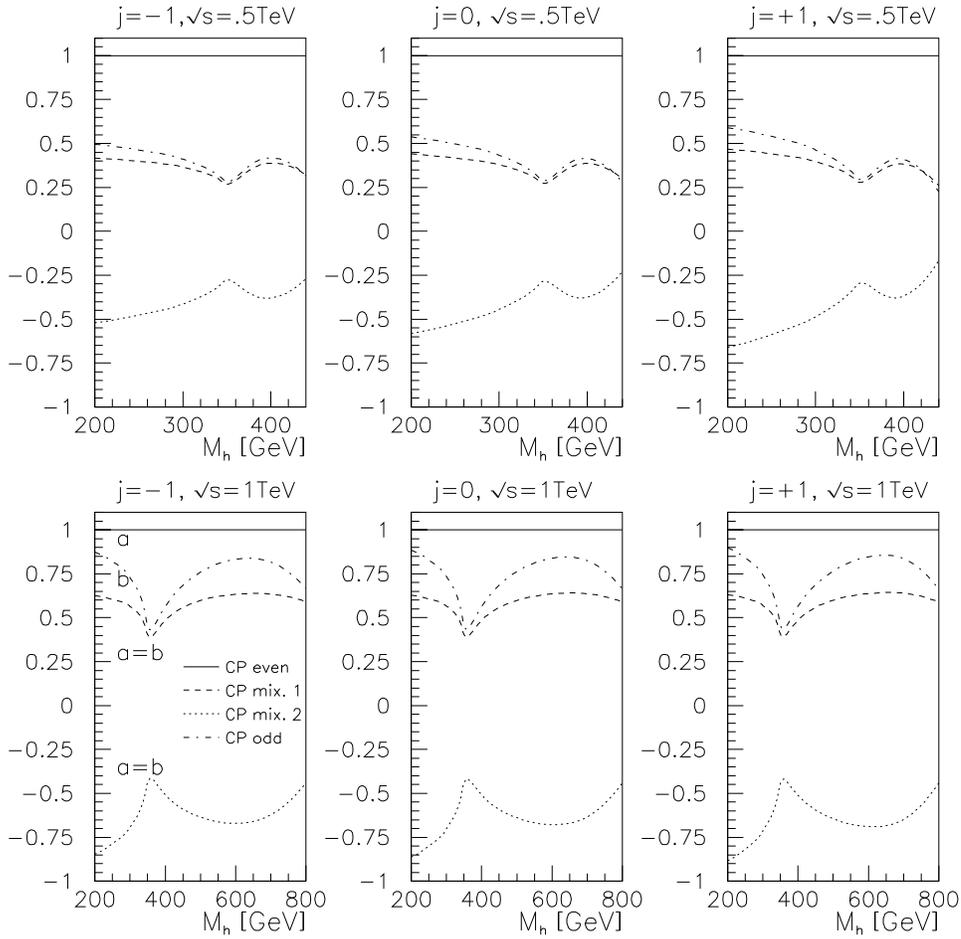}{.8}
\vspace*{-2.5cm}
\caption{Solutions for $a$ and $b$ as a function the the Higgs-boson mass.
For a given $M_h$, plotted $a$ and $b$ reproduce SM values for
$\Gamma_h$ and $\sigma_{tot}(\eettz)$ for various polarizations of the
initial electron beam.}
\label{abvalues}
\end{figure}

For given $M_H$ and $\sqrt{s}$ values, the width $\Gamma_h$ and the
cross section 
$\sigma_{tot}(\eettz)$ could be read directly from Fig.~\ref{cross}.
Specific values for $a$, $b$ and $c$ for the last three models are obtained by
solving the quadratic 
equations resulting from our requirements of $\Gamma_h=\Gamma_h^{SM}$ and
$\sigma_{tot}(\eettz)=\sigma_{tot}^{SM}(\eettz)$~\footnote{It should be
mentioned that relaxing this constraint may lead to different conclusions,
see Ref.~\cite{Atwood-Soni}, where models corresponding to very distinct
$\Gamma_h$ and $\sigma_{tot}(\eettz)$ have been compared.
We have checked that for models compared in Ref.~\cite{Atwood-Soni}
either widths or total cross sections differ by about a factor of $2$.}. 
As the equations to be solved are quadratic we have several solutions, 
we have chosen to discuss only those with the value of $c$ closest to $1$.
In the CP-mixed case there are both solutions with $a=b>0$ and $a=b<0$
we chose to discuss both cases.
Since for all considered
models it turns out that $c$ is very close to $1$ (for $M_h \leq
\sqrt{s}-\mz$ we have found $0.98 \leq c \leq 1.05$, for higher masses $c=1\pm.07$)
we present in 
Fig.~\ref{abvalues} only solutions for 
$a$ and $b$ as a function of $M_h$. As it is seen, substantial deviations
from the SM are possible keeping $\Gamma_h=\Gamma_h^{SM}$ and
$\sigma_{tot}(\eettz)=\sigma_{tot}^{SM}(\eettz)$. Since the total cross
section (see Fig.~\ref{cross}) depends on the initial beam
polarization the solutions for $a$ and $b$ also vary with the polarization.
Kinks in the curves correspond to the region where Higgs-boson
contribution to the total cross section for $\sigma_{tot}(\eettz)$ is
maximal, i.e. in the vicinity of the threshold for a real Higgs boson decaying
into $\ttbar$. 

\begin{figure}[h]
\vspace*{-1.5cm}
\postscript{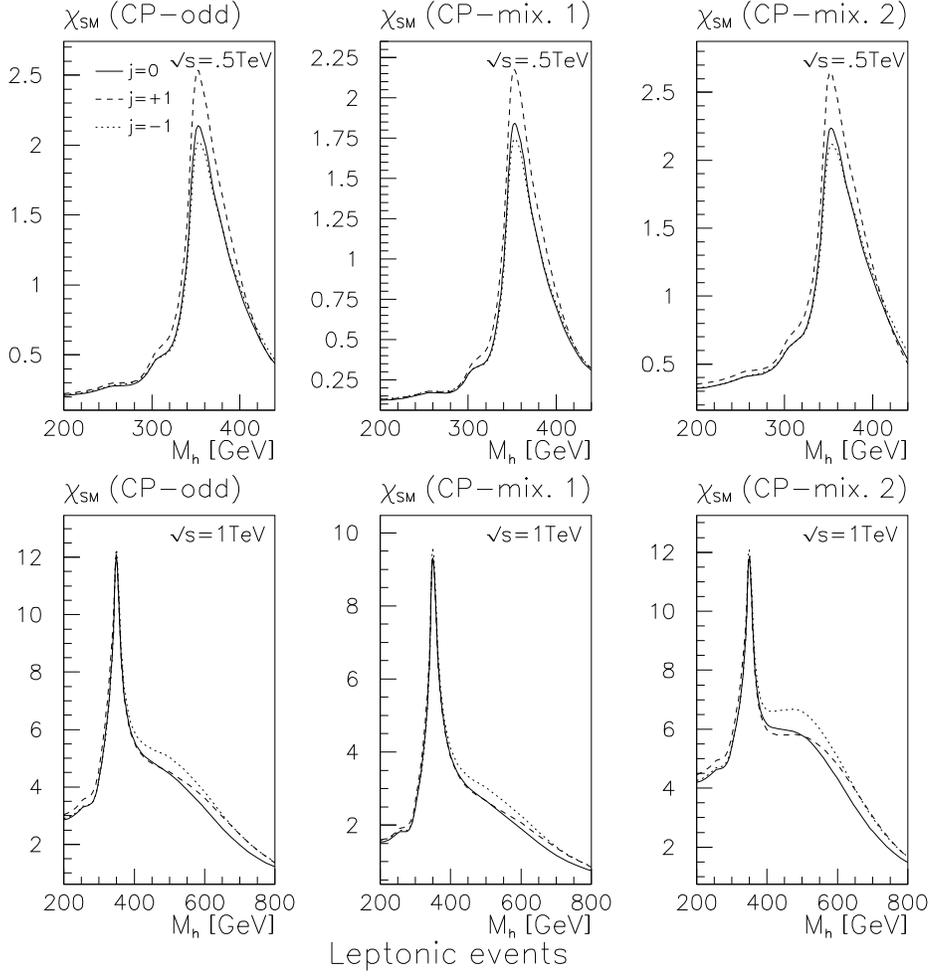}{.8}
\vspace*{-2.cm}
\caption{$\chi_X(Y) \equiv \protect\sqrt{\chi_X^2(Y)}$ where $\chi_X^2(Y)$ is
defined by Eq.({\protect\ref{chisqform}}) calculated for 
semileptonic top-quark decay final states (at least one $W$ boson decays
leptonically). The models considered are specified in the text. Results are
presented for polarized initial electron beams ($j=-1,0,+1$) and the total
energy $\protect\sqrt{s}=.5,\;1.\tev$.}
\label{chilepton}
\end{figure}

{\bf Since we stay within a class of models predicting the same}~\footnote{The
one measured in the future at LHC. Here it is assumed to be
$\Gamma_h^{SM}$.} {\boldmath $\Gamma_h$}{\bf, therefore we are entitled to
adopt the method 
of the optimal observables since the functions $f_{i}$ from
Eq.(\ref{sigform}) are known indeed}. In Figs.~\ref{chilepton} and
\ref{chihadron} we show results for $\chi_{X}(Y) \equiv
\sqrt{\chi_X^2(Y)}$ calculated according 
to Eq.(\ref{chisqform}) for the leptonic and hadronic events,
respectively. First of all we can see that for $\sqrt{s}=.5\tev$ 
even if the real Higgs boson is decaying into $\ttbar$,  $\chi$ never
exceeds $2.6$ both for leptonic and 
hadronic top-quark decays, making any
conclusions very uncertain. The initial electron polarization may be
helpful to some extend, 
although its effect is never very large. For $\sqrt{s}=1.\tev$ we
observe for the leptonic events that $\chi_{ SM}(CP-odd)$ is reaching 
$12.0$ just above $\mh=2\mt$ and it is above $3.8$ up to
$M_h=600\gev$. For the hadronic events the maximal $\chi$ is more or less
the same, however the region of large $\chi$ is narrower. 
This should be understood as an consequence of averaging
over $t$ and $\tbar$ as it suppresses contributions from terms
linear $bc$ leaving just the contribution from the square of the
Higgs-exchange. 

\begin{figure}[h]
\vspace*{-1.5cm}
\postscript{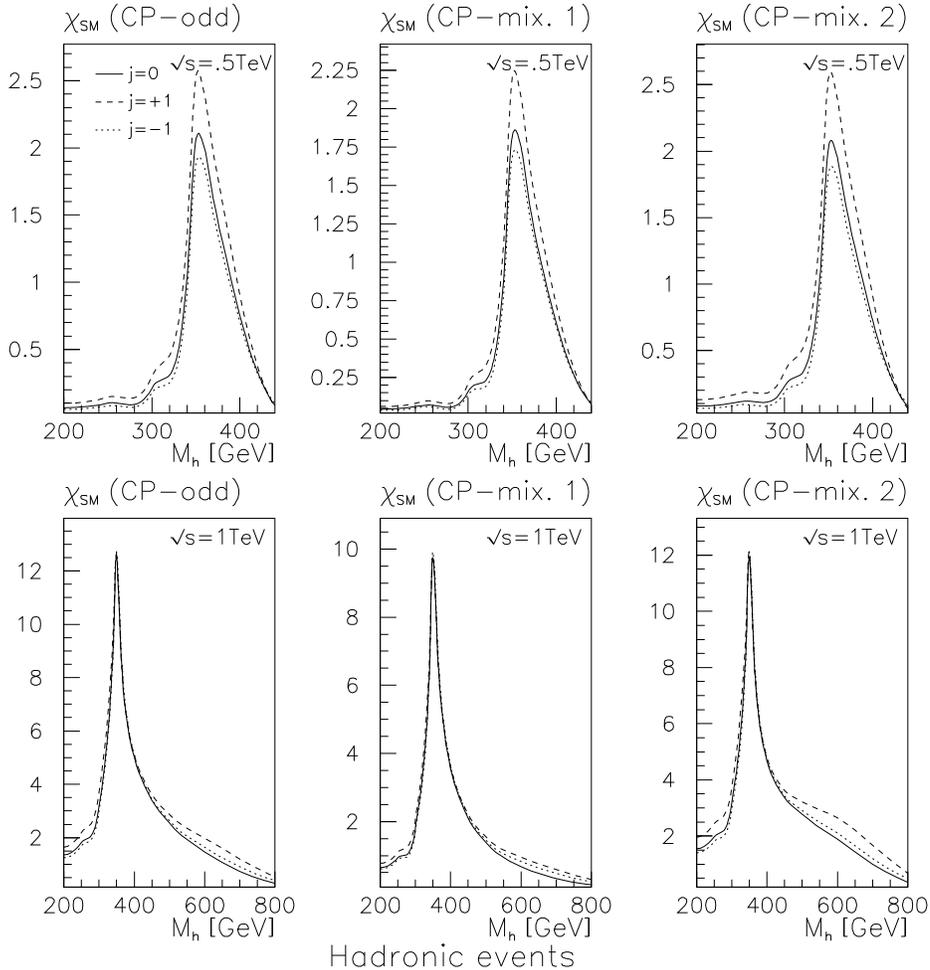}{.8}
\vspace*{-2.cm}
\caption{$\chi_X(Y)$ as in Fig.~{\protect\ref{chilepton}} but for purely hadronic
top-quark decays.}
\label{chihadron}
\end{figure}

It is worth mentioning that even though Fig.~\ref{interfer} was generated for
the SM it is, in part, still applicable in the context of the extensions defined
above. This 
is because the constraint $\Gamma_h=\Gamma_h^{SM}$  implies for $c \simeq 1$
that $\Gamma_h(h \ra \ttbar)$ should also be of the SM size. Therefore,
approximately, the contribution from the Higgs-boson diagram squared should
be close to the SM one for $M_h>2\mt$~\footnote{As it is saturated for
$M_h>2\mt$ by
the Higgs-boson resonance contribution :
$\sigma_{tot}(\eettz)\simeq\sigma_{tot}(hZ)\Gamma_h(h \ra \ttbar)/\Gamma_h$}. Since 
$\sigma_{tot}(\eettz)=\sigma_{tot}^{SM}(\eettz)$, it is obvious that the
interference between the Higgs and non-Higgs diagrams must also be the
same. 

One could notice that,
in spite of the fact that
$\sigma_{tot}(j=+1)<\sigma_{tot}(j=0)<\sigma_{tot}(j=-1)<$ 
there are such models that for certain Higgs-boson masses we have  
$\chi(j=+1)>\chi(j=0)>\chi(j=-1)$. 
One should however remember that for a given $M_h$ the model parameters
$a$, $b$ and $c$
adopted for $j=\pm,0$ are not the same as they are adjusted to reproduce
the SM prediction for $\Gamma_h$ and $\sigma_{tot}(\eettz)$, see
Fig~\ref{abvalues}. The functions $f_i$ utilized according to
Eq.(\ref{ciform}), for the calculation of 
$M_{ij}$ are also polarization dependent. In consequence it may happen that
the polarization leading to a lower total number of events provides higher
$\chi$ for certain models.

One could also try to disentangle the CP-odd model and the models of mixed-CP-Yukawa
couplings. However it turns out that the $\chi_{CP-odd}(CP-mix.1)$ is always below
$0.4$ and $2.3$ for $\sqrt{s}=.5\;{\rm and}\;1.\tev$, respectively. In other
words the models are too similar to be distinguished even by the method of
optimal observables. As for $\chi_{CP-odd}(CP-mix.2)$ the maximal values
obtained at $\sqrt{s}=.5\;{\rm and}\;1.\tev$ are $1.0$ and $6.2$,
respectively. Leptonic decay modes led always to higher $\chi$ while
distinguishing the CP-odd model and models of mixed-CP-Yukawa couplings. 

It should be mentioned here that if $\mh$ turned out to be just above
$2\mt$ then non-zero top-quark-width effects would smear out the ``spikes''
observed in Figs.~\ref{chilepton}, \ref{chihadron}, by an amount or the
order of the width $\Gamma_t \simeq 1.5\gev$. In that case one can expect
the QCD corrections to be relevant, of the order of 25\%~\footnote{For the
analogous process $\epem\ra\ttbar h$ the QCD corrections are
known~\cite{ditt} and for $\mh=100\gev\simeq\mz$ they have been found to be at
the level of 35\%.}.

So far we have been considering only production of unpolarized $\ttbar$
pairs. However, it is well known that helicity of the top quark could be
statistically determined through its semileptonic decay modes, e.g. lepton
energy distribution is sensitive to the initial top-quark helicity~\cite{bg-zh}.  
It is easy to notice that for polarized $\ttbar$ pairs differential
distributions would contain, in addition to $bc$, also other possible
signal of CP violation, namely terms proportional to $ab$. To perform a
realistic analysis one should consider $t$, $\tbar$ and $Z$ decays and then
having 8-body final phase space determine $ab$ and $bc$ applying the
method of optimal observables, this is however beyond the scope of this
paper. 

\section{Summary and Conclusions}

The method of the optimal observables~\cite{oo} has been applied to study
Yukawa couplings of the top quark in the process $\eettz$. The high
luminosity ($L=500\fbi$)  
option of linear $e^+e^-$ colliders examined in the context of the TESLA collider
design~\cite{tesla} has been adopted. Using the optimal weighting function
$\omega_{bc}$ given by Eq.(\ref{omega}) we have found that the minimal
statistical error for establishing CP violation in the scalar sector through
determination of $bc$, $\delta(bc)$ could be of the order of $.25$ for $M_h\simeq
450\gev$ for the $\sqrt{s}=1.\tev$ collider. The optimal observables have been
used to disentangle the SM (CP-even) and CP-odd or CP-mixed Yukawa
couplings. In order to make analysis more realistic only models predicting
the total Higgs-boson width and the total cross section for $\eettz$ equal
to those of the SM have been considered. It turned out that the lower
energy option $\sqrt{s}=.5\tev$ of 
the collider provides too small production rate to be applicable for the
determination of the top-quark Yukawa coupling. For $\sqrt{s}=1.\tev$
one could distinguish the models considered here
even at the level of $\chi=12$ (for $M_h \simeq 350 \gev$).
However, the results presented in this paper correspond to nearly ideal
experimental setup. The full analysis should include background and detector
simulation which would not only lower the efficiencies but also, due to the finite
detector resolution, would smear both the signal and background distributions. 
A detailed study including all those effects would be desired. 
Since the statistical significances found here were relatively large one
can believe that in a real experiment it would possible to determine the
top-quark Yukawa coupling in the region where $2\mt < \mh \lsim 600\gev$
for $\sqrt{s}=1\tev$.

\vspace{1.cm}
\centerline{\bf Acknowledgments}
\vspace{.5cm} This work was supported in part by the State Committee for
Scientific Research (Poland) under grant No. 2~P03B~014~14 and by Maria
Sklodowska-Curie Joint Fund II 
(Poland-USA) under grant No. MEN/NSF-96-252. The authors are indebted
to the U.C. Davis Institute for High Energy Physics for the great
hospitality extended to them while this work has being completed.

\vspace{.5cm}
\noindent{\Large\bf Appendix }
\vspace{.5cm}

Below we present contributions to the matrix element squared which are
linear in $bc$. 
As it is seen we have separated terms proportional to the real ($\re{\Pi_h}$)
and imaginary ($\im{\Pi_h}$) part of the Higgs-boson propagator
$\Pi_h(p_h^2)=[p_h^2-m_h^2+i m_h \Gamma_h]^{-1}$ for $p_h=p_t+p_{\tbar}$:
\begin{eqnarray}
|{\cal{M}}_{bc}|^2&=&bc \frac{g^6 m_t^2}{128 \cos^6\theta_W m_Z^2} 
\Pipropkmz \zvepjzae \\
&&\sum_{V=Z,\gamma} \Pipropkmzg \zgvepjzgae \left\{ R_V
\RePih + I_V \ImPih \right\}, \non
\end{eqnarray}
where 
\begin{eqnarray}
I_V&=&- \left\{ \right.
\Piqpqb \left[
-2j(\zat \zgat + \zvt \zgvt) \ft m_Z^2 s
-\zat \zgvt  (\et \ft - \gt \hti) (\hti - s)+ 
\right. \non\\
&&+\zvt \zgat [(\et \ft - \gt \hti) m_Z^2 + 
\gt (\ft^2 - (\hti - s)^2 + m_Z^2 s)]\left. \right]+ \non\\
&&\Piqmqb \left[j \zat \zgat \left[ 
\et[\hti^2 + (2 \hti - s) (m_Z^2 - s)]-\ft
\gt(\hti+2 m_Z^2)\right]+ \right.\non \\ 
&&\left.\left.
[\zgat \zvt (s-m_Z^2)+(\zat \zgvt-\zgat \zvt)\hti]
[\ft^2-(\hti-s)^2-m_Z^2 s]\right]\right\} \non \\
R_V&=&4 \mEpsilon \zat 
[\Piqpqb j \zgvt (\hti + \mzg^2 - s) - \Piqmqb \zgat \ft ]\non 
\end{eqnarray}
It was useful to adopt above the following CP-even and CP-odd quantities: 
\begin{eqnarray}
\et&=&(p_--p_+)\cdot(p_t-p_{\bar{t}})\non \\
\ft&=&(p_--p_+)\cdot(p_t+p_{\bar{t}})\\
\gt&=&(p_-+p_+)\cdot(p_t-p_{\bar{t}})\non \\
\hti&=&(p_-+p_+)\cdot(p_t+p_{\bar{t}})\non 
\end{eqnarray}
The vector-boson propagators are defined as $\Pi_V(s)=[s-m_V^2]^{-1}$.
For quark propagators the following notation has been used:
\beq
\Piqpqb=\Pi_t\left[(k+p_t)^2\right]+
\Pi_{\bar{t}}\left[(k+p_{\bar{t}})^2\right]\;\;\;
\Piqmqb=\Pi_t\left[(k+p_t)^2\right]-\Pi_{\bar{t}}\left[(k+p_{\bar{t}})^2\right],
\eeq
where $\Pi_x(q^2)=[q^2-m_x^2]^{-1}$.  

\begin{table}[h]
\bce
\begin{tabular}{|c|c|c|}
\hline
$V$&$\zgvf$&$\zgaf$\\
\hline
$Z$&$2 T_3^f-4 Q^f \stwsq$&$-2 T_3^f$\\
\hline
$\gamma$&$4 Q^f \stw\ctw $&$0$\\
\hline
\end{tabular}
\hspace{2cm}
\begin{tabular}{|c|c|c|}
\hline
$f$&$T_3^f$&$Q^f$\\
\hline
$e$&$-\frac{1}{2}$&$-1$\\
\hline
$t$&$+\frac{1}{2}$&$+\frac{2}{3}$\\
\hline
\end{tabular}
\ece
\caption{The vector and axial-vector couplings
of the electron and of the top quark.}
\end{table}
\vspace{1cm}

The contribution proportional to the real part of the Higgs propagator have
been already published in the literature~\cite{Atwood-Soni}. We have
noticed two missprints present in the published result~\cite{Atwood-Soni}:
factor $1/2$ is 
missing in Eq.(12) and sign in front of the first term in Eq.(13) should be
reversed. 

It has been argued in Ref.~\cite{Atwood-Soni} that the contribution
from the imaginary part of the Higgs propagator should be
omitted since it is of higher order as being proportional to the Higgs
width $\Gamma_h$. However, one should have in mind that the integration
over the phase 
space around the Higgs-boson mass provides an extra factor $\Gamma_h^{-1}$
which cancels $\Gamma_h$ from the numerator of the imaginary part of the propagator leading
to the contribution of the same order as the one proportional to the real
part. This is exactly the same mechanism which is
responsible for the applicability of the narrow width approximation.
However, to generate the compensating factor $\Gamma_h^{-1}$ it is necessary
to have the pole of the Higgs-boson propagator with its vicinity (of the
size of the $\Gamma_h$) within the integration region. 
In the process under consideration, the square of the Higgs boson momentum,
$p_h^2=(p_t+p_{\tbar})^2$ is 
limited by $4\mts$ and $s-\mzs$. It is easy to find that the
position in the middle of the allowed region corresponds to $m_h=430$ and
$750\gev$ for $\sqrt{s}=.5$ and $1\tev$, respectively. Since both for
$430\gev$ and $750\gev$ 
the allowed region contains the Higgs-boson pole with its vicinity of the
size of the width one can expect
contributions of the same order 
from both real and imaginary parts of the Higgs-boson propagator. 
We have found that the effect of the imaginary part could increase
$\chi_{CP-odd}(CP-mix.1)$ even by $200\%$, however for that case $\chi$
was still 
below $1$. The only phenomenologically relevant case for which the imaginary
part was important was $\chi_{SM}(CP-mix.2)$ for $M_h \simeq 500\gev$
where we have found $10\%$ increase caused by the imaginary part.
It turns out that for all other considered cases the contribution 
from the imaginary part of the Higgs-boson propagator could be neglected for
all practical purposes, although a consistent
treatment demands both the real and imaginary contributions.

\end{document}